\documentclass[preprint,3p]{elsarticle}
\usepackage{amssymb}
\usepackage{soul}

\usepackage[utf8]{inputenc} 
\usepackage[T1]{fontenc}    
\usepackage{hyperref}       
\usepackage{url}            
\usepackage{booktabs}       
\usepackage{amsfonts}       
\usepackage{nicefrac}       
\usepackage{microtype}      
\usepackage{lipsum}		
\usepackage{graphicx}
\usepackage{natbib}
\usepackage{doi}

\usepackage{bm}        
\usepackage{amssymb}   
\usepackage{amsmath}   
\usepackage{color} 
\usepackage{lineno}
\usepackage{multirow}
\usepackage{enumerate}
\usepackage{comment}
\usepackage{tabularx}
\usepackage{booktabs}
\usepackage{array}
\usepackage{soul}

\usepackage{caption}
\usepackage{subcaption}

\newcolumntype{P}[1]{>{\centering\arraybackslash}p{#1}}
\newcolumntype{Y}{>{\centering\arraybackslash}X}

\usepackage{gensymb}

\def\Mpc{\operatorname{Mpc}}

\def\lapp{\ifmmode\stackrel{<}{_{\sim}}\else$\stackrel{<}{_{\sim}}$\fi}
\def\gapp{\ifmmode\stackrel{>}{_{\sim}}\else$\stackrel{<}{_{\sim}}$\fi}

\usepackage{dcolumn}
\usepackage{bm}

\usepackage{caption}
\usepackage{subcaption}

\newcolumntype{P}[1]{>{\centering\arraybackslash}p{#1}}
\newcolumntype{Y}{>{\centering\arraybackslash}X}

\usepackage{gensymb}

\def\Mpc{\operatorname{Mpc}}

\def\lapp{\ifmmode\stackrel{<}{_{\sim}}\else$\stackrel{<}{_{\sim}}$\fi}
\def\gapp{\ifmmode\stackrel{>}{_{\sim}}\else$\stackrel{<}{_{\sim}}$\fi}

\usepackage{caption}
\journal{.}

\begin{document}

\begin{frontmatter}



\title{Gravitational wave signals from long lasting binary--single black hole encounters.}


\author[a,b]{Elena Codazzo}
\author[c,d]{Matteo Di Giovanni} 
\author[a,b]{Jan Harms}

\affiliation[a]{Gran Sasso Science Institute (GSSI), I-67100 L`Aquila, Italy}
\affiliation[b]{INFN, Laboratori Nazionali del Gran Sasso, I-67100 Assergi, Italy}
\affiliation[c]{Dipartimento di Fisica, Universita` di Roma La Sapienza, I-00185 Rome, Italy}
\affiliation[d]{INFN, Sezione di Roma, I-00185 Rome, Italy}

\begin{abstract}
In the dense regions of star clusters, close encounters with black holes (BHs)  can occur giving rise to a new class of gravitational-wave (GW) signals. Binary-single encounters between three BHs are expected to dominate the rate of signals from unbound systems in the frequency band of terrestrial GW detectors. The encounter can describe a quasi-hyperbolic trajectory, which was the focus of a recent study. In some cases, the encounter can take a more complex form including one or two BH mergers as a result of the encounter, repeating cycles of close proximity between the BHs, and the exchange of a BH that is part of the binary. The variety of types of encounters leads to a variety of GW signals emerging from these encounters. Using the ARWV N-body code, we performed 42 numerical simulations, to explore various outcomes of binary-single interaction, and we characterize the diverse GW signatures produced during these encounters. Additionally, we evaluated the detectability of these GW signals by injecting them into the simulated noise of the Einstein Telescope and exploring different methods to detect the signals. Our findings shed light on the complexities of these interactions and their potential implications for GW astronomy.
\end{abstract}




\end{frontmatter}


\section{Introduction \label{introduction}}

After the first successful observation runs with the current-generation gravitational-wave (GW) detectors Advanced Virgo \citep{aVirgo} and Advanced LIGO \citep{aLIGO}, the GW community is preparing to leap into the 3rd generation of GW detectors, such as the proposed Einstein Telescope (ET) \citep{et,ET2020} and Cosmic Explorer \citep{EvEA2021}. The improved sensitivity of these future ground-based detectors is expected to increase the number of observed events from tens per year to hundreds of thousands per year opening an enormous science case \citep{MaEA2020,KaEA2021a}. Moreover, the addition of the space-based detector LISA \citep{ASEA2017} together with pulsar-timing arrays \citep{HoEA2010} and possibly with decihertz Moon-based GW detectors \citep{gloc, lgwa} will enable the scientific community to cover the GW spectrum from nanohertz to kilohertz.

Taking all these elements into consideration, the accessible frequency band of the GW spectrum will be significantly broadened, therefore prompting a series of studies focused at the GW emission from non-canonical sources, i.e., different from compact object mergers, continuous waves and a stochastic background, that are out of reach for current detectors.

Among these sources are hyperbolic encounters between compact objects, black holes (BHs) in particular, which are enduring a growing interest in the astrophysical and GW communities after the detection of GW190521 \citep{gw190521det, abbott2020properties}, the most massive BH merger detected so far. Given its masses and eccentricity of the orbit, some follow-up studies \citep{dallamicoGW190521, gayathri2022eccentricity, Gamba:2021gap} suggested that the merger could have been the outcome of the dynamical interaction between two unbound solar-mass BHs, therefore providing the first indirect detection of dynamical interaction in a star cluster. Previously, the GW170817A binary black-hole (BBH) merger had been found consistent with a hierarchical merger due to dynamical interactions \citep{Gayathri_2020} as well, but the large uncertainties in the reconstructed parameters damped its significance. Later, also the GW190425 \citep{gw190425} binary neutron star (BNS) merger, given its unusually high total mass, led to the hypothesis, as one of the possible formation scenarios, of a dynamical encounter in a cluster between a BNS and an isolated neutron star that replaced the lightest component of the binary. Although this explanation appears intriguing, it has been shown that the dynamical formation channel has a negligible contribution to the BNS merger rate in the local universe, therefore making the dynamical origin for GW190425 unlikely \citep{refId0, Ye_2020}. Nevertheless, the binary pulsars B2127$+$11C in M15 \citep{anderson} and J1807$-$2500 in NGC 6544 \citep{Lynch_2012} still provide evidence of neutron stars that experienced dynamical interactions in clusters \citep{Andrews_2019}.

Dynamical interactions are not new to the GW community. Since the start of the century, several theoretical studies have focused on the analytical \citep{8capozziello2008gravitational,majar2010gravitational,9de2012gravitational,de2014gravitational,cho2018gravitational,99garcia2018gravitational,mitra2021detectability,morras2022search} and numerical \citep{damour2014strong,nagar} tools to determine the GW emission of a fly-by or of a dynamical capture. Recently, linking for the first time the astrophysical properties about star population and cluster dynamics \citep{portegies2000,banerjee2010,tanikawa2013,oleary, mapellietal2013, ziosietal,rodriguez2015,rodriguez2016,rodriguez2019, mapelli2016,askar2017, seddaetal,samsing2018,samsing2018b,fragione2018,fragione2019,fragione2022,zevin2019,zevin2021,kremer2019, Sedda_2020, mapellietal2021, Rastello,banerjee2021,rizzuto2022,kamlah2022} with the GW emission of hyperbolic flybys, \citep{Codazzo23} provided a first realistic estimate of the detectability of gravitational radiation from binary-single hyperbolic flyby between BHs in star clusters and the rate of these events.

As a follow up to \citep{Codazzo23}, the aim of this work is to identify, within a set of injections in simulated ET noise, which are the typical features of these signals. This should enable us to determine, among other things, not only if there are characteristics in common that could facilitate their recognition by a data-analysis pipeline, but also whether a typical eccentric merger, characterized by an inspiral phase, was preceded by a dynamical interaction or represents a second-generation merger following an initial capture event.

The paper is organized as follows: in Section \ref{background}, an overview of the underlying astrophysical assumptions of the systems considered in this study is provided; Section \ref{methods} details the simulations carried out for this work and the methodology used to simulate BBH-BH encounters and to characterize their GW signals. The results are presented and discussed in Section \ref{results}. A short summary along with final remarks conclude the paper in Section \ref{conclusions}.

\section{Astrophysical background}\label{background}
Nuclear star clusters (NSC) are one of the most promising locations for dynamical encounters between compact objects \citep{heggie1975binary, hutbahcall}. Since an extensive overview of NSC and binary-single encounters has already been given in \citep{Codazzo23} and other past works \citep{tremaine1975formation,capuzzo1993evolution,antonini2012dissipationless,pfeffer2018mosaics, hopkins2010nuclear,hopkins2010massive,mapelli2012situ, guillard2016new,dall2023eccentric}, here we will limit ourselves to the highlights of their properties.

In star clusters, binary systems are classified as \textit{hard} if their binding energy is higher than the average kinetic energy of the stars in the cluster \citep{heggie1975binary}. BBHs in the core harden not only for GW emission but also through binary--single encounters \citep{heggie1975binary}, namely three-body interactions in the form of a fly-by or a resonant interaction between the BBH and an intruder object. In the latter case, an unstable triple system is formed until one of the three objects acquires enough speed to escape; typically, it is the least massive component that most probably escapes \citep{1980AJ.....85.1281H}.
In general, flybys are the most common outcome of such interactions. 
Since, statistically, the velocity of the intruder after the encounter is greater than the one with which it approached the binary, as a consequence of the conservation of energy, the binary tightens according to Heggie's law, for which hard binaries tend to become harder.

The semi-major axis \textit{a} of the binary will decrease over time due to binary--single encounters and due to GW emission \citep{peters1964gravitational} as follows \citep{Mapelli2020}:
\begin{equation}\label{sma_hardening}
\dfrac{\text{d}a}{\text{d}t}= -2 \pi \xi\dfrac{G \rho_{\rm c}}{\sigma} a^2 -\frac{64}{5} \dfrac{G^3 m_1 m_2(m_1+m_2)}{c^5a^3(1-e^2)^{7/2}}f_1(e).
\end{equation}
where $m_1$ and $m_2$ are the primary and the secondary mass of the BBH, \textit{e} is its eccentricity, $m_3$ is the mass of the intruder, $\sigma$ is the three-dimensional dispersion velocity, $ \rho_{\rm c} $ is the local density of stars, $\xi\approx $ 3 is a dimensionless hardening rate \citep{quinlan1996dynamical} and
\begin{equation}
f_1(e)=\left( 1+\frac{73}{24}e^2 + \frac{37}{96}e^4 \right).
\end{equation}
The contribution of dynamical hardening, proportional to $a^2$, dominates over that due to the GW emission by the system, which instead takes over when the semi-major axis is small since it is proportional to $a^{-3} $ \citep{peters1964gravitational}.

Similarly, dynamical encounters and GW emission \cite{peters1964gravitational} cause the binary eccentricity to change over time as \citep{Mapelli2020}:
\begin{equation}
\dfrac{\text{d}e}{\text{d}t}= 2 \pi \xi \kappa \dfrac{G \rho_{\rm c}}{\sigma} a -\frac{304}{15} e \dfrac{G^3 m_1 m_2(m_1+m_2)}{c^5a^4(1-e^2)^{5/2}}f_2(e).
\end{equation}
with the $\kappa$ parameter defined as $\kappa \equiv \text{d}e / \text{d ln}(1/a) $\cite{quinlan1990dynamical};
and
\begin{equation}
f_2(e)=\left( 1+\frac{121}{304}e^2\right).
\end{equation}

Using these properties, in \citep{Codazzo23}, we conducted a study on the detectability of gravitational signals from binary-single hyperbolic encounters. We used stellar-mass black holes as input masses and considered hard binaries in an environment that is the core of an NSC.

Our findings indicate that the rate of these encounters within NSCs, up to redshift z\,=\,3.5, falls within the range of $[0.006 - 0.345]\,$yr$^{-1}\,$Gpc$^{-3}$. For the lower and upper limits, we considered core radii of 0.1\,pc and 1\,pc, respectively.
We observed that the majority of these encounters produced GW signals in the frequency band of LISA, but no signals were in the ground-based interferometer frequency band. However, we discovered a correlation between the characteristic frequency of the signals with the binary's semi-major axis and the impact parameter with which the intruder approaches. In this study, we tuned these parameters to obtain signals suitable for analysis within the frequency band of ET. All the parameters considered are aligned with the presence of a binary and an intruder in an environment resembling the core of an NSC, which we extensively examined in our previous work \citep{Codazzo23}.

\section{Method and simulations}\label{methods}
To simulate binary-single encounters, we use the N-body simulation code ARWV \citep{mikkola1993implementation, chassonnery2, chassonnery1}. ARWV makes use of the algorithmic regularization chain method to integrate the equations of motion \cite{mikkolaEOM1989} \cite{mikkolaEOM1993}. In this way the round-off errors are reduced, making the regularization algorithm more efficient,
especially for close interactions. ARWV incorporates a post-Newtonian treatment up to order 2.5 to correct the equations of motion in case of strong gravitational interactions \cite{mikkolaPN2008}. These simulations provide the positions of all three bodies at each time step, allowing for computation of the strength of the gravitational signal emitted by their configuration.

The initial conditions are chosen to represent realistic encounters within the core of a NSC, as already described by \citep{Codazzo23}. Furthermore, \citep{Codazzo23} found that the frequency of the GW signal increases as both the binary's semi-major axis and the impact parameter of the intruder decrease. Therefore, after the generation of the initial conditions, we select only the systems with impact parameter and semi-major axis eligible to generate a GW signal within the ET frequency band and sensitivity. Finally, we use the Hilbert-Huang Transform analysis \citep{Huang1998} to reveal distinctive features of the signals.

\subsection{Initial conditions}\label{initial_conditions}
The simulation of a binary-single encounter requires the use of a set of 15 parameters to define the initial conditions of the simulation \cite{hutbahcall,dall2021gw190521,Codazzo23} including the spins of the BHs. Among these parameters are the angles $\phi$, $\psi$, $\theta$ that define the geometry of the encounter, the impact parameter $b$, the velocities of the BHs and the orbital parameters of the binary. A detailed description of these parameters is beyond the scope of this paper, but can be found in \cite{hutbahcall,dall2021gw190521,Codazzo23}. All these parameters are sampled according to the distributions described in \citep{Codazzo23}. Moreover, from the initial conditions generated for \citep{Codazzo23}, we select an event with initial conditions eligible to generate a GW signal detectable by ET (Table \ref{table_e0}).

Using this event as a reference, we decided to divide the simulations into two categories: spinning BHs and non-spinning BHs. In the former case, the spins are randomly sampled according to a Maxwellian distribution with a root mean square of 0.1. For each category, we redefine the reference event as {\it e0} (Table \ref{table_e0}) for spinless simulations and {\it e5} for simulations with spinning BHs. The initial conditions within the two categories are exactly the same, with the only difference being the spin values. At this point, we decide to gradually move from {\it e0} and {\it e5} by changing only one parameter at a time while keeping all the others fixed in each simulation. The goal of this procedure is to assess whether the spin has an impact on the outcome of the encounters or not and to show how small and predictable changes in the initial conditions can lead to unpredictable outputs that dramatically affect the outcome and the detectability of each event.

In particular, we decide to keep the spins and the geometry of the encounter fixed for all simulations and to change only the following parameters:
\begin{itemize}
\item eccentricity: e $\in$ [0, 0.5, 1, 1.5, 2] $\times 10^{-5}$;
\item relative velocity between binary and intruder: v $\in$ [30, 40, 50, 60, 70] km/s;
\item impact parameter: b $\in$ [3, 4, 5, 6, 7] $\times 10^{-3}$ AU;
\item intruder mass: m$_3 \in$ [5, 7, 10, 13, 15] M$\odot$;
\item mass ratio of the binary: q $\in$ [0.3, 0.5, 0.7, 0.83, 1].
\end{itemize}
In total, we ran 42 simulations, equally divided between the spinning and non-spinning BHs cases.

\begin{table}[h]
\centering
\captionsetup{width=\textwidth}
\centering
\caption{Initial values of the parameters for the baseline event {\it e0}. From top to bottom: the two masses of the binary, the mass of the intruder, the semi-major axis and the eccentricity of the binary, relative velocity and impact parameter of the intruder approaching the binary, and the geometrical parameters (fase, $\psi$, $\phi$, and $\theta$) fixed for all the simulations. All three BHs spinless.}
\begin{tabular}{c|c}
\hline
Parameters  &  Initial Values \\

\hline

m$_1$ [M$_\odot$]  & 30.0 \\
m$_2$ [M$_\odot$]  & 25.0 \\
m$_3$ [M$_\odot$]  & 10.0 \\
a [AU] & 8e-05 \\
e & 0.0 \\
v [km/s] & 50.0 \\
b [AU] & 0.005 \\
fase & 2.8 \\
$\psi$ & 3.0 \\
$\phi$ & 3.0 \\
$\theta$ & 1.5 \\
spins & 0.0 \\

\hline
\end{tabular}

\label{table_e0}
\end{table}

\subsection{Characterization of the signal}\label{characterization}
\begin{figure*}[h!]
\includegraphics[width=\textwidth]{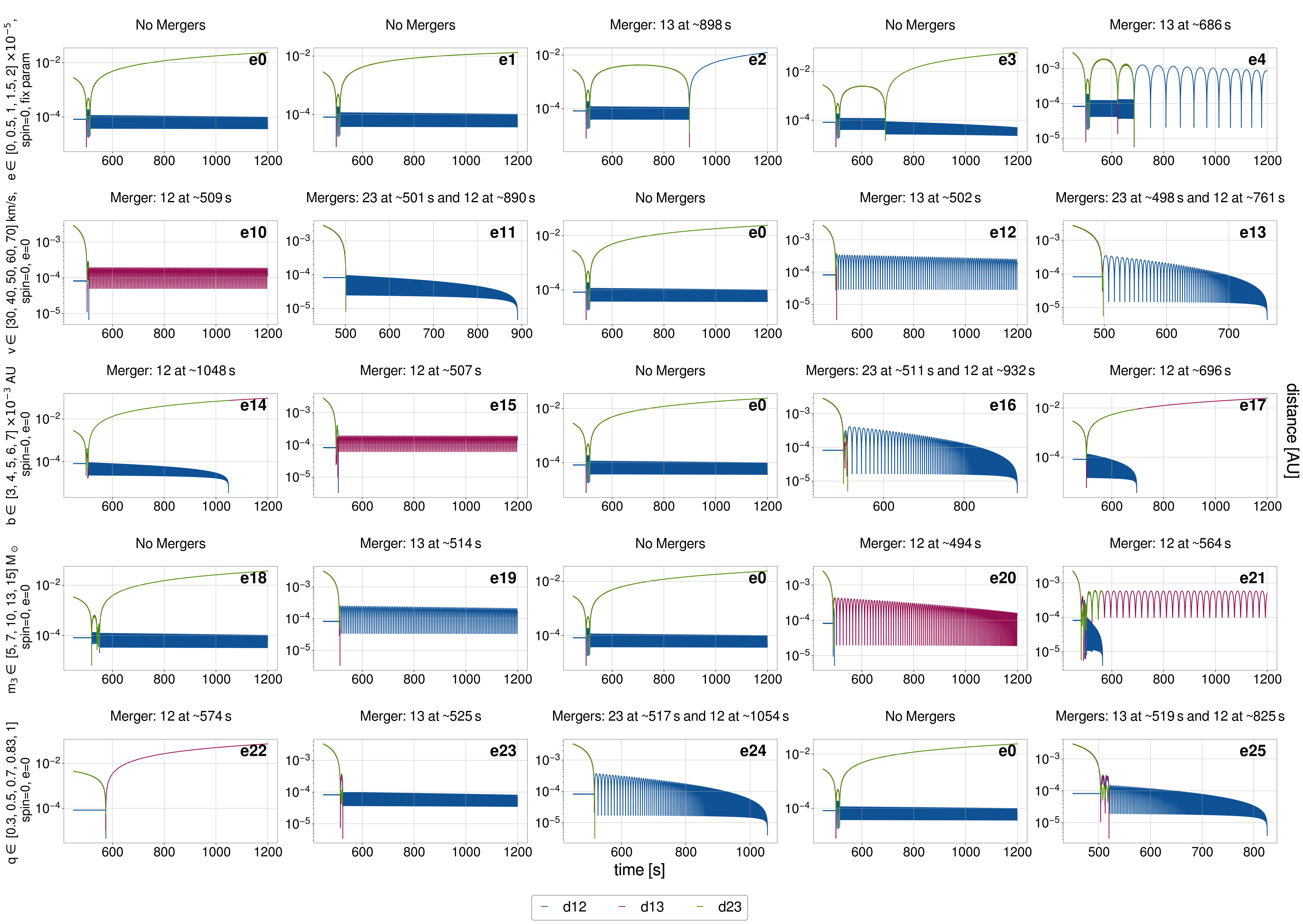}
\caption{Mutual distance among the three bodies as a function of time for simulations with initial BH spins set to zero. Each plot in the 5x5 grid displays a unique simulation. Each row represents the variation of a single parameter in the initial conditions, in particular eccentricity e $\in$ [0, 0.5, 1, 1.5, 2] $\times 10^{-5}$ (first row);
relative velocity between binary and intruder v $\in$ [30, 40, 50, 60, 70] km/s (second row); impact parameter b $\in$ [3, 4, 5, 6, 7] $\times 10^{-3}$ AU (third row); intruder mass m$3 \in$ [5, 7, 10, 13, 15] M$\odot$ (fourth row); mass ratio q $\in$ [0.3, 0.5, 0.7, 0.83, 1] (bottom row). The title of each panel highlights the occurrence of merging bodies, their identity in parentheses, along with the time of the merger.
The event {\it e0} serves as baseline for all events and is shown multiple times for comparison.}
\label{distances_nonspin}
\centering
\end{figure*}

\begin{figure*}[h!]
\includegraphics[width=\textwidth]{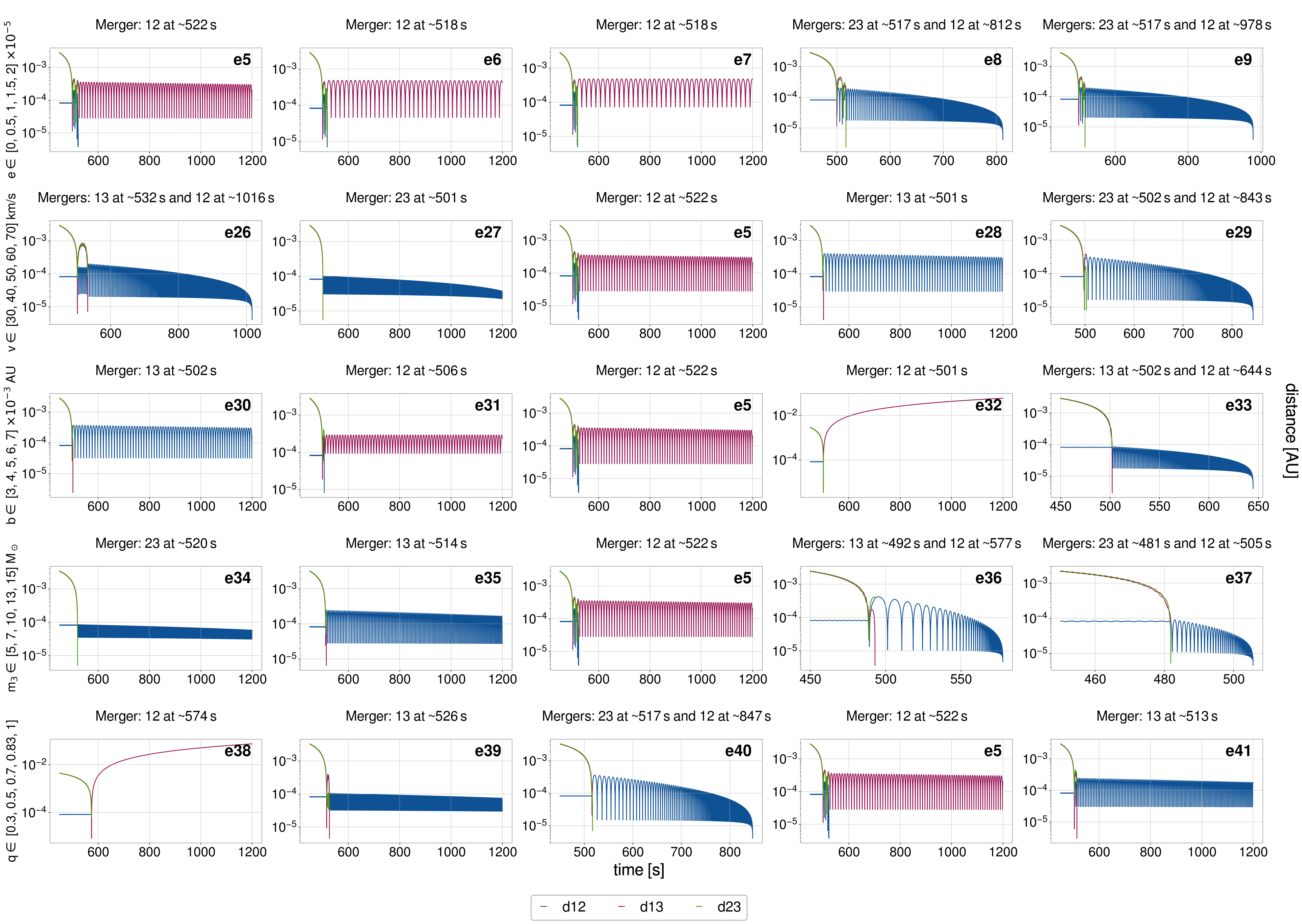}
\caption{Same as Figure \ref{distances_nonspin}, but in all simulations the BHs have initial spins. The reference event is denoted as {\it e5}.}
\label{distances_spin}
\centering
\end{figure*}

Through the output of ARWV, we track the evolution of each BBH-BH encounter. Figures \ref{distances_nonspin} and \ref{distances_spin} show the time evolution of the mutual distances among the three bodies for non-spinning and spinning BHs respectively, for different initial conditions. The baseline events {\it e0} and {\it e5} are also depicted once for each row for comparison and the outcome of the simulations will be discussed in Sec. \ref{results}.

To characterize the GW emission of each event, we compute the wave amplitude as \cite{ferrari2020general}:
\begin{equation}
h^{TT}_{jk}=\dfrac{2G}{c^4 R} \ddot{Q}^{\rm TT}_{jk},  \label{hTT}
\end{equation}
where $j,\,k=$1,2 and 3, \textit{R} is the luminosity distance of the source from us and $\ddot{Q}^{\rm TT}_{jk}$ is the second time derivative of the transverse-traceless part of the quadrupole moment. We set our sources at a distance of 230$\Mpc$. From now on, we consider the wave propagating along the $z-$direction and, of the full GW tensor, we choose the $h_{11}$ component only, i.e. the plus polarization h$_+$, as GW amplitude of our signals.

\begin{figure*}[h!]
\centering
\includegraphics[width=0.9\textwidth]{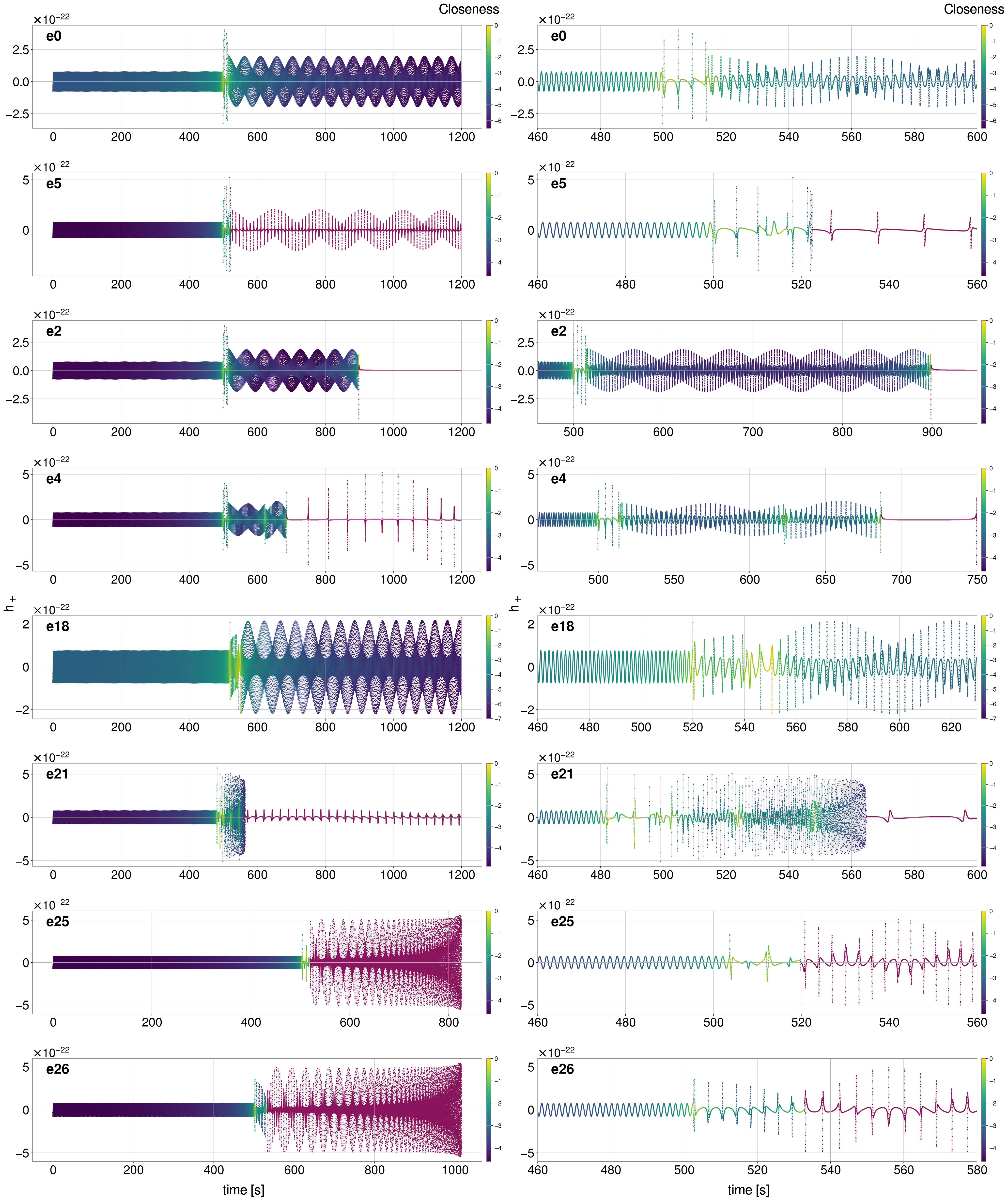}
\caption{{\it Left}. The plots illustrate eight time-series, each representing the signal of an entire simulation - denoted on the top left. Each simulation lasts 1200 seconds unless two mergers occur, leading to an early termination. The color scale corresponds to the closeness parameter, with the magenta color denoting a value of zero, indicating the presence of only two remaining bodies due to the merger of two out of three bodies. {\it Right}. The plots display zoomed-in views of the corresponding plots on the left, focusing on encounters and mergers, providing detailed insights into the evolution of the signal and closeness parameter during critical phases of the simulation.}
\label{CLOSENESS}
\centering
\end{figure*}

In order to investigate the influence of the third body on the signal, we define a parameter, referred to as \textit{closeness}, that provides a measure of their spatial distance as:
\begin{equation}  
C = \text{log}  \frac{\text{d}_{min}}{\text{d}_{max}},
\end{equation}
where d$_{min}$ and d$_{max}$ are the minimum and the maximum values of the three mutual distances and $C<0$ by definition. Figure \ref{CLOSENESS} presents the time evolution of h$_+$ during 1200\,s long simulations, unless they are interrupted before. Interruptions occur when, due to consecutive mergers, only one of the three bodies is left in the simulation. The closeness parameter is visualized using a color scale with magenta identifying $C=0$, which specifies that the number of bodies in the system decreased from three to two due to a merger.

\subsubsection{Hilbert-Huang Transform}\label{hht}
We use the Hilbert Huang Transform (HHT) \citep{Huang1998} to characterize the signals, utilizing the \texttt{EMD} Python Package \citep{Quinn2021_joss}. HHT is a signal processing technique designed for the analysis of non-linear and non-stationary data. This method involves two steps. First, it employs empirical mode decomposition (EMD) to decompose complex signals into intrinsic mode functions (IMF) and residuals, effectively segregating the time domain data into various frequency scales. These IMFs, when aggregated with the residual, reproduce the original time series. Notably, this sifting process is performed adaptively, how the modes are biased depend on the data towards locally dominant frequencies.
Subsequently, the HHT method applies the Hilbert Transform to each IMF, facilitating the computation of instantaneous frequency (IF) and amplitude (IA) for each IMF with respect to time. This capability allows for the monitoring of temporal fluctuations in the frequency and amplitude of signal components.

What the Hilbert transform actually does is give an analytical signal, {\it z(t)}, from real data, {\it f(t)}, to calculate instantaneous properties.
The analytical signal will be:
\begin{equation}
z(t)=f(t)+i H\{f(t)\}=A(t) e^{i \theta(t)}.
\end{equation}
$H\{f(t)\}$ is the Hilbert transform of  $f(t)$ :
\begin{equation}
H\{f(t)\}=\frac{1}{\pi} \mathcal{P} \int_{-\infty}^{\infty} \frac{f\left(\tau\right)}{t-\tau} d \tau,
\end{equation}
where $\mathcal{P}$ indicates the Cauchy principal value.
$A(t)$ and $\theta(t)$ are the amplitude and the phase of the signal defined as:
\begin{equation}
\begin{aligned}
A(t)&=\sqrt{f^2(t)+H\{f(t)\}^2}, \\
\theta(t)&=\tan ^{-1}[H\{f(t)\} / f(t)].
\end{aligned}
\end{equation}
The IF and IA to compute the Hilbert spectrum are respectively:
\begin{equation}
\begin{aligned}
\omega(t)&=\frac{d \theta(t)}{d t}, \\
e&=|A(t)|^2.
\end{aligned}
\end{equation}

\subsection{Signal injections in ET noise}\label{Injection}
\begin{figure*}[h!]
\centering
\includegraphics[width=0.8\textwidth]{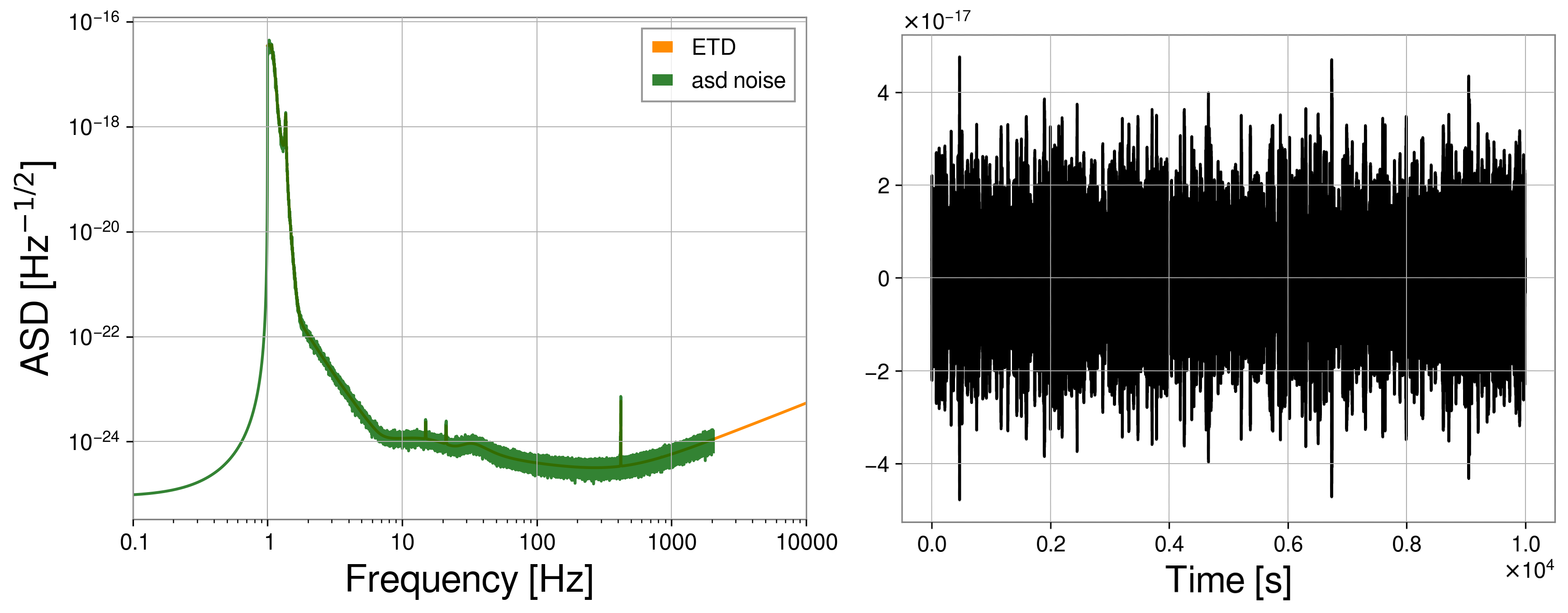}
\caption{{\it Left side}: amplitude spectral density of the ET simulated noise in green compared to its sensitivity curve \citep{hild2011sensitivity}. {\it Right side}: 10$^4$\,s of ET simulated noise in the time domain. } \label{noise}
\end{figure*}

The signals obtained from our simulations are injected into 10$^4$\,s long data segments of ET noise sampled at 4096\,Hz. ET noise is simulated by generating white Gaussian noise colored with the ET design sensitivity curve \citep{hild2011sensitivity}. We use the Welch to obtain the amplitude spectral density (ASD) of the noise. In doing so, we divide the data into segments of 10$^3$\,s in length, with an
overlap of 0.5. The ASD of the noise is shown in Figure \ref{noise} together with the same noise in the time domain. The injections are done through the \texttt{gwpy} \citep{gwpy, duncan_macleod_2023_7821575} and \texttt{pycbc} \citep{alex_nitz_2023_7692098} Python packages.

We use the ASD of the signals, $S_h$, and the ASD of the noise, $S_n$, to calculate the signal-to-noise ratio (SNR) as follows:
\begin{equation}\label{SNR_formula}
SNR = \sqrt{4 \cdot \sum_{k=1}^{N} \left( \frac{S_h(f_k)^2}{S_n(f_k)^2} \right)}
\end{equation}
The SNR is related to the area between the characteristic strain $h_c$ and the characteristic noise strain $h_n$, computed as:
\begin{equation}\label{3b+gw}
\begin{aligned}
h_c&=f\sqrt{T S_h} , \\
h_n&=\sqrt{f S_n},
\end{aligned}
\end{equation}
where {\it T} is the segment length. The plots of the characteristic strain of all the simulated signals are shown in Figures \ref{Cstrain_ns} and \ref{Cstrain_s} of the Appendix.

The time series of the signals injected into the noise were tapered and bandpassed between 2\,Hz and 30\,Hz to compute the spectrograms of these data, lasting up to 1200\,s, containing the signals.

\section{Results}\label{results}
In the first instance, we discuss the outcome of the simulated encounters, where the components of the binaries are identified as bodies 1 and 2 whereas the intruder is body 3. Just considering the two baseline events {\it e0} and {\it e5}, identical except for the spins, the distinction between the non-spinning (Figure \ref{distances_nonspin}) and spinning case (Figure \ref{distances_spin}) significantly alters the outcome of the two events. In the former case, as can be better seen from the time series in Figure \ref{CLOSENESS}, the third body completes two close orbits around the binary, which remains gravitationally bound even after the encounter, resulting in an increase of its eccentricity. In contrast, in the latter case, the third body interacts multiple times with the binary and leads it to a merger. The newly formed BH then forms a new binary system with the third body, remaining gravitationally bound for the rest of the simulation, with an eccentricity of approximately 0.8.
In events {\it e2} and {\it e4} (Figure \ref{distances_nonspin} and \ref{CLOSENESS}), after the initial encounter, the system remains bound for several orbits until the merger of bodies 1 and 3 occurs. In {\it e2}, the resulting BH and body 2 move away from each other, while in {\it e4}, a new binary with high eccentricity forms.
The event {\it e21} (Figure \ref{distances_nonspin} and \ref{CLOSENESS}) exhibits a particularly intriguing behavior. Following the encounter, the third body remains gravitationally bound to the binary, which, due to the interaction, enters the inspiral phase. A close examination of the time series reveals that the third body undergoes several close fly-bys during the inspiral phase, remaining bound to the new BH in a binary system with an eccentricity of 0.7. In {\it e25} and {\it e26} (Figure \ref{distances_spin} and \ref{CLOSENESS}), the simulations are interrupted shortly after, 800\,s and 1000\,s respectively,  due to a second merger. While we expect similar mergers to occur when a second-generation binary is formed, we are not able to notice them in our simulations since this could happen after the time limit of 1200\,s. What is important to note here is that only the second merger occurs through the spiraling of the two bodies, while in both cases, the first merger is a rapid plunge resulting from the close interaction.

To understand the features of the stronger events, we characterize the high SNR regime in the ideal case in which the SNR is calculated using the characteristic strain $h_c$ of the simulated signal over the noise strain $h_n$ of simulated ET noise. After that, we inject the signal in the ET noise to have an insight on what to expect in the real case and whether the insights obtained from our simulations can be useful to analyze the real signal or not.

\begin{figure*}[h!]
\includegraphics[width=\textwidth]{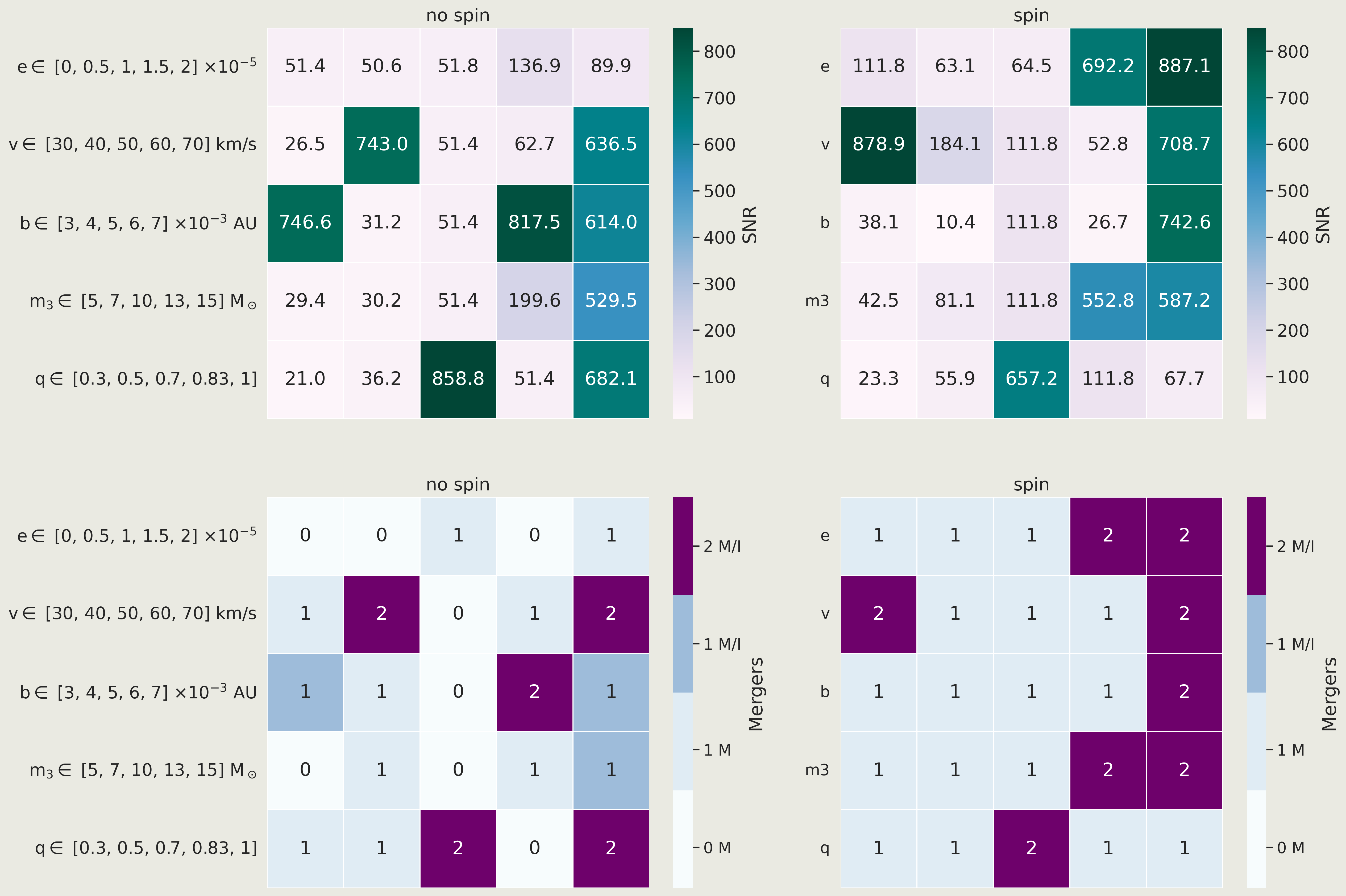}
\caption{{\it Top}. Heatmaps showing the SNRs for Figures \ref{distances_nonspin} (left) and \ref{distances_spin} (right). {\it Bottom}. Heatmaps illustrate the number of mergers occurring in the simulations. The arrangement of the 5x5 grid corresponds to Figures \ref{distances_nonspin} (left) and \ref{distances_spin} (right). The two heatmaps exhibit discrete color variations representing different scenarios: zero mergers (0\,M), one merger resulting from plunge or capture events (1\,M), one merger following an inspiral between two bodies (1\,M/I), and two mergers, with at least the second involving an inspiral phase (2\,M/I). Notably, the presence of the inspiral phase significantly increases the SNR, as evident when comparing the cells in the heatmaps with 1\,M/I and 2\,M/I to the corresponding SNR heatmap cells. The presence of an inspiral phase, even if interrupted by the end of the simulation, leads to an increase in the SNR, even in cases with zero mergers (e.g., {\it e3}) or with one merger (e.g., {\it e5}, {\it e20}, {\it e27}, {\it e35}).}
\label{heatmap}
\centering
\end{figure*}

The top row of Figure \ref{heatmap} shows the SNR of the simulated signals, whereas the bottom row shows the corresponding number of mergers per event. The mergers are categorized as follows: 
\begin{itemize}
    \item no mergers: \textit{0\,M};
    \item one merger resulting from a plunge or capture event: \textit{1\,M};
    \item one merger following an inspiral phase between two bodies: \textit{1\,M/I};
    \item two mergers, with at least the second one involving an inspiral phase: \textit{2\,M/I}.
\end{itemize}

It becomes apparent that the presence of an inspiral leads to a higher SNR, as clearly seen by comparing the heatmap showing the number of mergers with the SNR heatmap in Figure \ref{heatmap}. Of particular interest are the cases with 0\,M (e.g., {\it e3}) or 1\,M (e.g., {\it e5}, {\it e20}, {\it e27}, {\it e35}), where unexpectedly high SNRs for the type of event are observed. This is due to the presence of an inspiral that has been cut by the end of the simulation because the merger time was longer than the simulation time. The striking observation from Figure \ref{heatmap} is that spinning systems invariably lead to at least one merger event, unlike their non spinning counterparts.

\begin{figure*}[h!]
\centering
\includegraphics[width=0.8\textwidth]{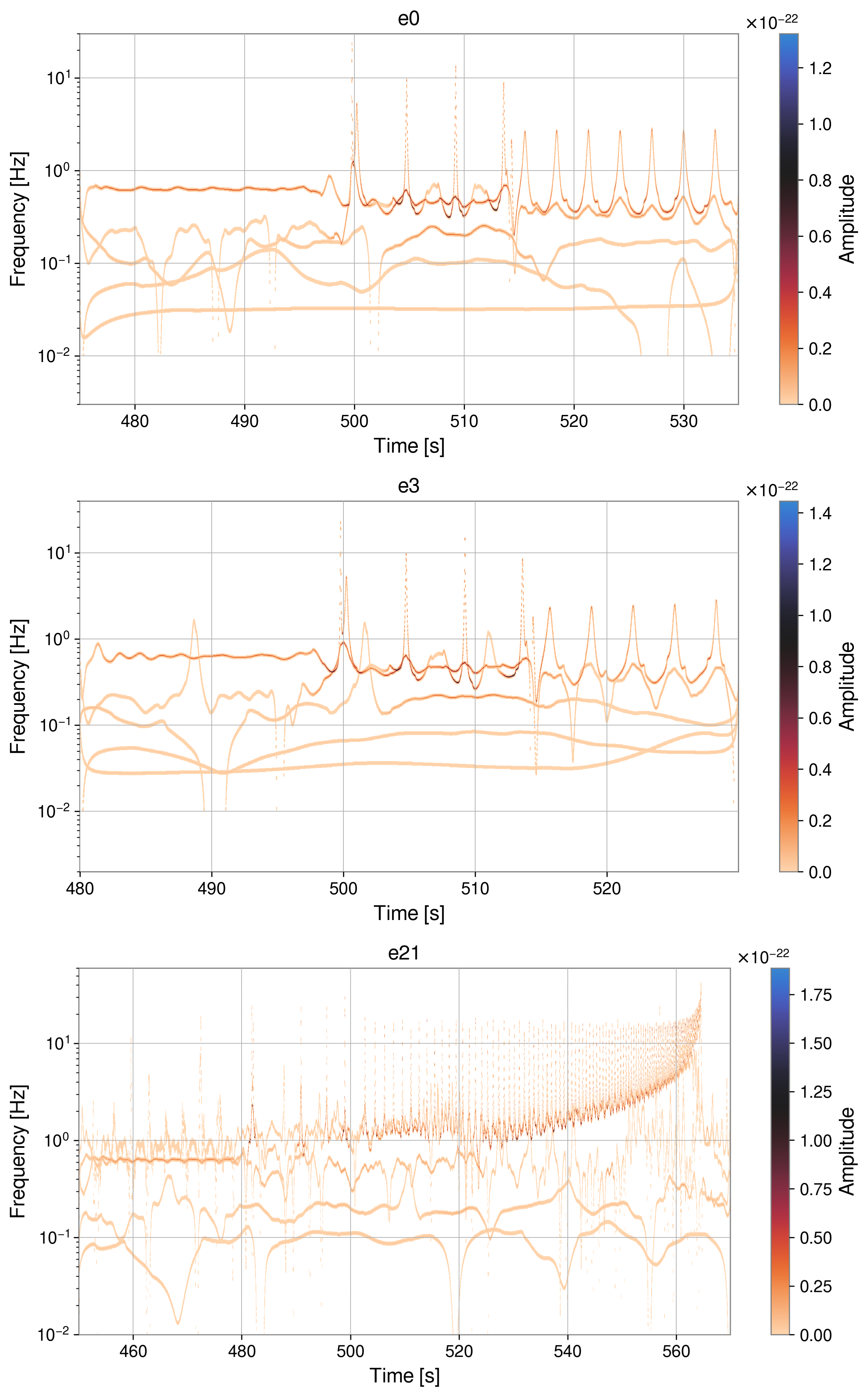}
\caption{HHT spectrum of simulations {\it e0}, {\it e3} and {\it e21}. Encounters cause rapid changes in frequency. } \label{hht_spec}
\end{figure*}

We then compute the HHT of the simulated signal across time to explore changes in patterns of the instantaneous frequency. Figure \ref{hht_spec} shows the HHT of a segment of the simulations {\it e0}, {\it e3} and {\it e21}. 
As we can see for {\it e0}, from $T = 500\,s$ the IMF containing the highest frequencies shows sharp peaks. In the higher IMFs, IMF\,1 and IMF\,2, the trace of the two encounters, happening at 500\,s and 515\,s respectively (Figure \ref{CLOSENESS} and \ref{hht_spec}), is particularly visible. Before the encounter, the binary with $e\sim0$ keeps a steady value of IMF\,1; after the first encounter, IMF\,1 changes by showing sharp frequency peaks; after the second encounter the peaks change again. This is also visible by the rapid increase of IMF\,2. All in all, the close encounters between the intruder and the binary, which drastically modify the semi-major axis and the eccentricity of the binary, show sudden frequency changes in the HHT. A very similar pattern can be observed in the plot referring to event {\it e3}, with a first steady frequency trace, followed by sudden peaks reaching up to 10\,Hz and by a further change after the second encounter. A slightly different pattern can be observed for {\it e21}. During this resonant encounter, we have multiple fly-bys - at 485\,s, 490\,s, 505\,s, 525\,s and 545\,s - due to the formation of a triple system. The inspiraling motion triggered after the encounter at 505\,s translates to a more fuzzy frequency pattern. The following encounters are still visible both in the time series and in the HHT: in particular, we can see that, after the inspiral starts at 505\,s, another encounter at 525\,s produces a clearly visible change in the IMF trend. 

\begin{figure*}[h!]
\centering
\includegraphics[width=0.4\textwidth]{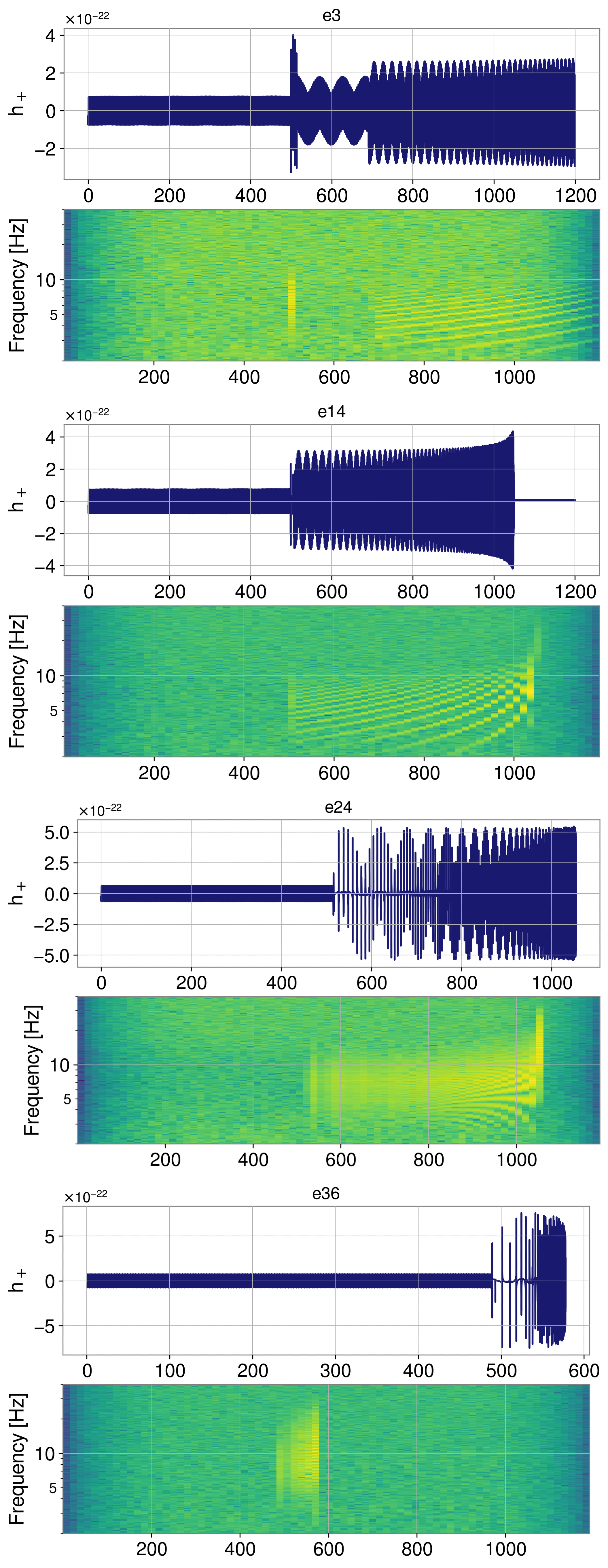}
\caption{Time series and spectrograms of four simulations: {\it e3}, {\it e14}, {\it e24} and {\it e36}. The top panel of each event contains the h$_+$ as a function of time for the entire simulation. The spectrograms of the signals injected in the noise - tapered and bandpassed - are normalized by the median of the frequency bins.} \label{spectrogram}
\end{figure*}
Finally, we proceed to inject the signals in simulated ET noise. The plots in Figure \ref{spectrogram} show the spectrograms, each 1200\,s long, of {\it e3}, {\it e14}, {\it e24} and {\it e36}. The data, sampled at 4096\,Hz, are tapered and bandpassed between 2\,Hz and 30\,Hz and then apply the Welch method using 16\,s long segments with an 8\,s overlap to compute the spectra.
Each frequency bin is normalized by its median to highlight the time variation.
\clearpage
There is a notable excess of power during the first encounter, as shown in the examples in Figure \ref{spectrogram}, a signature evident in all 42 simulations. Due to the encounters, the binary systems exhibit high eccentricity, a distinctive feature visible in the spectrograms of {\it e3} and {\it e14} with multiple tails during the inspiral. 
In simulation {\it e3}, there are three encounters of interest. The first two are very close to each other, at around the 500\,s, whereas the third is just before 700\,s. The third encounter leads to the inspiral of the binary system. Despite the simulation ending before the actual merger, the phase preceding it is clearly visible in the spectrogram. Simulation {\it e14} exhibits a similar behavior, where the initial two encounters, characterized by an excess of power, promptly trigger the merger of the binary system, perfectly recognizable in the time-frequency plot.
In simulation {\it e24}, the initial interaction is a plunge. The intruder immediately merges with the less massive component of the binary system. However, the binary system itself remains bound with high eccentricity, approximately 0.9. The formation of this more massive and high eccentric binary is evident in the spectrogram, where after the first excess of power, unstable peaks manifest before the system gradually start to circularize.
Looking at the spectrogram of {\it e36}, it is clear that there is a gap between an initial burst (at 485\,s) and the subsequent merger (starting at 550\,s) with the inspiral part that follows, showing how another scenario - the burst - triggered the merger. Specifically, this burst that we can observe in the spectrogram refers to the close encounter and to the subsequent plunge. 
Although the specifics of the phenomenon underlying the burst cannot be discerned from the spectrograms alone, observing similar behavior in a significant percentage of our simulations suggests that the bursts preceding the mergers have an astrophysical nature and are indicative of a dynamic environment.

\section{Conclusions}\label{conclusions}

We simulated 42 single-binary encounters with the N-body code ARWV starting from a common set of initial conditions. For each simulation, we varied only one of these parameters, in order to show how sensitive the simulations are even to a small change in initial conditions. We then inspected the gravitational signals of these events, which involve mergers with their typical inspiral phase, as well as bursts due to encounters or plunges. We employed the HHT, a method used to analyze data exhibiting non-linear and non-stationary patterns. It decomposes signals into their individual components, enabling the analysis of their time-evolving frequency and amplitude characteristics. By applying this method, we were able to see clearly how encounters, after modifying some of the geometric parameters of the binary (semi-major axis and eccentricity), produce sudden and visible frequency changes. Finally, after the injection of our signals in the simulated ET noise, we were still able to observe these frequency changes in the spectrograms of the events.

Our findings suggest that a highly eccentric merger could be preceded by encounters - or even by previous mergers (plunges) - that act as boosters for the merger. In our experiments we considered very tight initial binaries, thus obtaining quite short simulations, lasting even less than 20 minutes; in reality, this behavior could also vary in a wider temporal range. By inspecting the spectrograms of the signals injected in simulated ET noise, we showed that the main features of these signals are distinguishable also without any prior knowledge on the nature of the event under examination.

For a future development of this work, we foresee the improvement of the simulation of spinning bodies. In fact, although ARWV includes the spins, it does not provide enough information about their orientation. This means that we are limited to the specific case in which the spins are assumed to be aligned and that we do not know how an encounter affects the orientation of the spins in a binary system. 

The meaningfulness of this development resides in the fact that, as we have already extensively discussed, dynamical encounters lead to some typical signatures like eccentric orbits, unusual mass ratios and misaligned spins. Therefore, the study of how dynamical interactions affect the spins of a binary system would be of great help to support follow-up studies of future detections, numerical waveform studies and detection pipeline developments. This will become paramount when, in future detectors, we will be able to detect also the continuous GW signal from the coalescence of compact objects from which we will be able to infer relevant information about the orbital parameters of the system.

\bibliographystyle{unsrtnat} 

\section{Appendix}

In this appendix, we report the Table (\ref{table_AllEvents}) containing the initial values of the parameters and the plots of the characteristic strains of all the simulations.

In Table \ref{table_AllEvents} are shown only the parameters that were modified in each simulation. As explained in Sec \ref{initial_conditions} the value that we varied from \textit{e0} or \textit{e5} is one between: eccentricity (\textit{e}), relative velocity (\textit{v}), impact parameter (\textit{b}), intruder mass (\textit{m$_3$}) and mass ratio of the binary (\textit{q}=m$_2$/m$_1$). 
In the simulations that include spins, the spins of the three BHs are taken from a Maxwellian distribution with a root mean square of 0.1 and do not change across the different simulations.

\begin{figure*}[t]
\includegraphics[width=\textwidth]{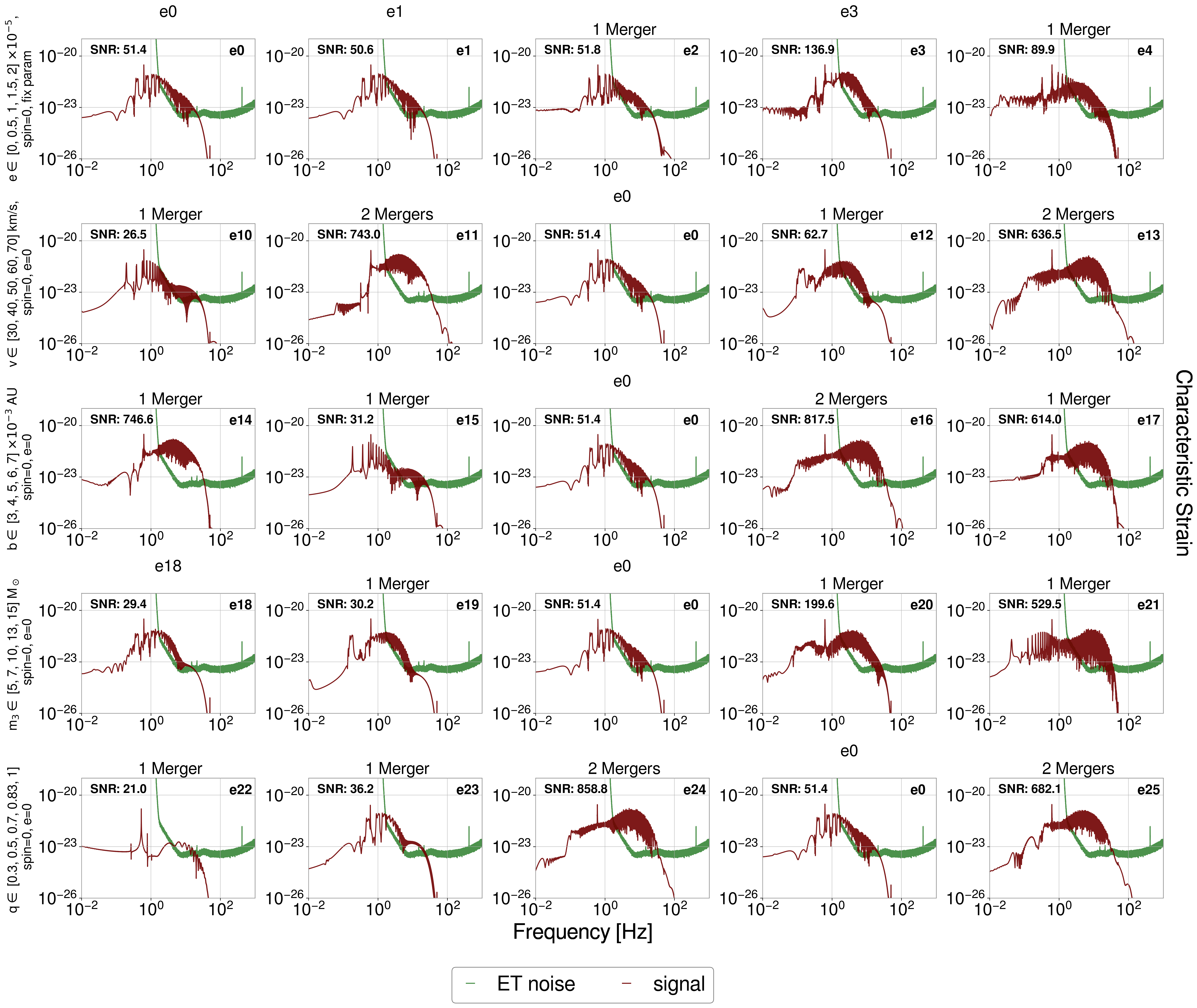}
\caption{characteristic strain (brown) of the gravitational signal from simulations in Figure \ref{distances_nonspin} and the strain amplitude of the ET noise (green). At the top of each plot, the simulation label is displayed. Furthermore, the SNR of each event is provided in the top-left corner of the corresponding plot.}
\label{Cstrain_ns}
\centering
\end{figure*}

\begin{figure*}[t]
\includegraphics[width=\textwidth]{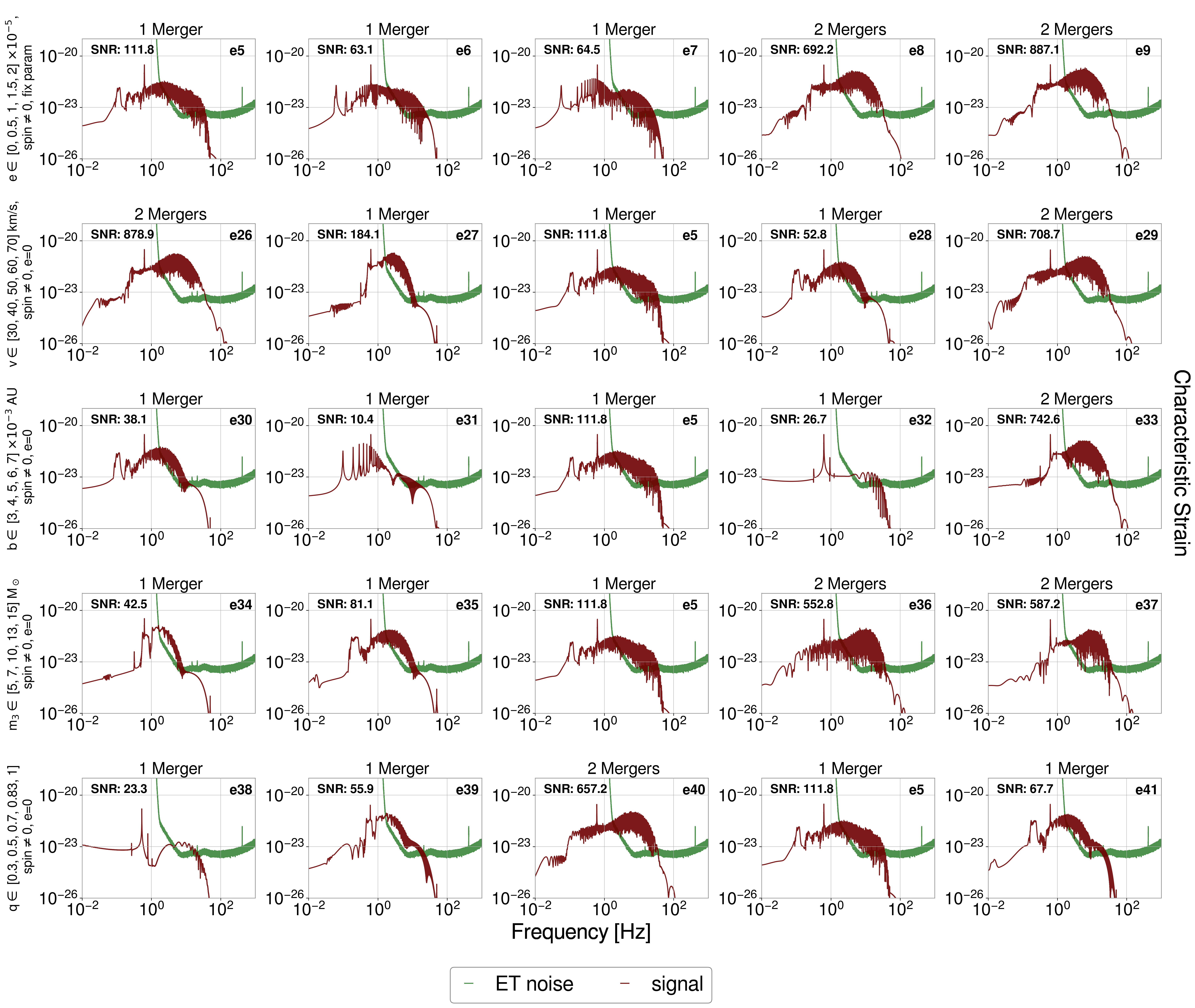}
\caption{Similar to Figure \ref{Cstrain_ns}, but displaying the characteristic strain of the signals from Figure \ref{distances_spin}.}
\label{Cstrain_s}
\centering
\end{figure*}

Figures \ref{Cstrain_ns} and \ref{Cstrain_s} illustrate the characteristic strain of the gravitational signal from simulations in Figure \ref{distances_nonspin} and \ref{distances_spin} respectively. The presence of a spike at 0.6\,Hz is attributed to the accumulated signal from the initial binary before the encounter. 

\begin{table}[h!]
\centering
\caption{Initial parameter values for all conducted simulations. Only values that may differ from simulation \textit{e0} are listed.}
\label{table_AllEvents}
\begin{tabular}{cccccccc}
\hline
event & m1 [M$_\odot$] & m2 [M$_\odot$] & m3 [M$_\odot$] & e & v [km/s] & b [AU]& spin \\
\hline
e0 & 30.0 & 25.0 & 10.0 & 0.0 & 50.0 & 0.005 & 0.0 \\
e1 & 30.0 & 25.0 & 10.0 & 5e-06 & 50.0 & 0.005 & 0.0 \\
e2 & 30.0 & 25.0 & 10.0 & 1e-05 & 50.0 & 0.005 & 0.0 \\
e3 & 30.0 & 25.0 & 10.0 & 1.5e-05 & 50.0 & 0.005 & 0.0 \\
e4 & 30.0 & 25.0 & 10.0 & 2e-05 & 50.0 & 0.005 & 0.0 \\
e5 & 30.0 & 25.0 & 10.0 & 0.0 & 50.0 & 0.005 & $\neq$ 0 \\
e6 & 30.0 & 25.0 & 10.0 & 5e-06 & 50.0 & 0.005 & $\neq$ 0 \\
e7 & 30.0 & 25.0 & 10.0 & 1e-05 & 50.0 & 0.005 & $\neq$ 0 \\
e8 & 30.0 & 25.0 & 10.0 & 1.5e-05 & 50.0 & 0.005 & $\neq$ 0 \\
e9 & 30.0 & 25.0 & 10.0 & 2e-05 & 50.0 & 0.005 & $\neq$ 0 \\
e10 & 30.0 & 25.0 & 10.0 & 0.0 & 30.0 & 0.005 & 0.0 \\
e11 & 30.0 & 25.0 & 10.0 & 0.0 & 40.0 & 0.005 & 0.0 \\
e12 & 30.0 & 25.0 & 10.0 & 0.0 & 60.0 & 0.005 & 0.0 \\
e13 & 30.0 & 25.0 & 10.0 & 0.0 & 70.0 & 0.005 & 0.0 \\
e14 & 30.0 & 25.0 & 10.0 & 0.0 & 50.0 & 0.003 & 0.0 \\
e15 & 30.0 & 25.0 & 10.0 & 0.0 & 50.0 & 0.004 & 0.0 \\
e16 & 30.0 & 25.0 & 10.0 & 0.0 & 50.0 & 0.006 & 0.0 \\
e17 & 30.0 & 25.0 & 10.0 & 0.0 & 50.0 & 0.007 & 0.0 \\
e18 & 30.0 & 25.0 & 5.0 & 0.0 & 50.0 & 0.005 & 0.0 \\
e19 & 30.0 & 25.0 & 7.0 & 0.0 & 50.0 & 0.005 & 0.0 \\
e20 & 30.0 & 25.0 & 13.0 & 0.0 & 50.0 & 0.005 & 0.0 \\
e21 & 30.0 & 25.0 & 15.0 & 0.0 & 50.0 & 0.005 & 0.0 \\
e22 & 30.0 & 9.0 & 10.0 & 0.0 & 50.0 & 0.005 & 0.0 \\
e23 & 34.0 & 17.0 & 10.0 & 0.0 & 50.0 & 0.005 & 0.0 \\
e24 & 30.0 & 21.0 & 10.0 & 0.0 & 50.0 & 0.005 & 0.0 \\
e25 & 27.0 & 27.0 & 10.0 & 0.0 & 50.0 & 0.005 & 0.0 \\
e26 & 30.0 & 25.0 & 10.0 & 0.0 & 30.0 & 0.005 & $\neq$ 0 \\
e27 & 30.0 & 25.0 & 10.0 & 0.0 & 40.0 & 0.005 & $\neq$ 0 \\
e28 & 30.0 & 25.0 & 10.0 & 0.0 & 60.0 & 0.005 & $\neq$ 0 \\
e29 & 30.0 & 25.0 & 10.0 & 0.0 & 70.0 & 0.005 & $\neq$ 0 \\
e30 & 30.0 & 25.0 & 10.0 & 0.0 & 50.0 & 0.003 & $\neq$ 0 \\
e31 & 30.0 & 25.0 & 10.0 & 0.0 & 50.0 & 0.004 & $\neq$ 0 \\
e32 & 30.0 & 25.0 & 10.0 & 0.0 & 50.0 & 0.006 & $\neq$ 0 \\
e33 & 30.0 & 25.0 & 10.0 & 0.0 & 50.0 & 0.007 & $\neq$ 0 \\
e34 & 30.0 & 25.0 & 5.0 & 0.0 & 50.0 & 0.005 & $\neq$ 0 \\
e35 & 30.0 & 25.0 & 7.0 & 0.0 & 50.0 & 0.005 & $\neq$ 0 \\
e36 & 30.0 & 25.0 & 13.0 & 0.0 & 50.0 & 0.005 & $\neq$ 0 \\
e37 & 30.0 & 25.0 & 15.0 & 0.0 & 50.0 & 0.005 & $\neq$ 0 \\
e38 & 30.0 & 9.0 & 10.0 & 0.0 & 50.0 & 0.005 & $\neq$ 0 \\
e39 & 34.0 & 17.0 & 10.0 & 0.0 & 50.0 & 0.005 & $\neq$ 0 \\
e40 & 30.0 & 21.0 & 10.0 & 0.0 & 50.0 & 0.005 & $\neq$ 0 \\
e41 & 27.0 & 27.0 & 10.0 & 0.0 & 50.0 & 0.005 & $\neq$ 0 \\
\hline
\end{tabular}
\end{table}


\end{document}